\documentclass[11pt,a4paper]{JHEP3}
\usepackage{amsmath, amssymb}
\usepackage{euscript}
\def\be{\begin{eqnarray}}
\def\ee{\end{eqnarray}}
\def\0{\nonumber}
\def\d{\partial}

\usepackage{amsfonts}
\usepackage{verbatim}
\newcommand{\CR}{\\\nonumber}

\newcommand{\refb}[1]{(\ref{#1})}

\newcommand{\VEV}[1]{\ensuremath{\,\langle{}#1\rangle\,}}

\newcommand\qqd{\quad\,\quad}

\def\k{\kappa}
\def\la{\lambda}
\def\u{{\tilde u}}

\newlength{\somewidth}
\newcommand{\nn}[1]{\settowidth{\somewidth}{\ensuremath{-1}}\makebox[\somewidth]{\ensuremath{#1}}}

\preprint{SISSA/38/2009/EP\\ZTF-09-01\\\tt hep-th/xxxx.yyyy}

\title{Hawking fluxes, Fermionic currents, $W_{1+\infty}$ algebra and anomalies}

\author{ L.Bonora$^a$, M.Cvitan$^{a,b}$, S.Pallua $^{b}$,
I.Smoli\'c $^b$\\
 $^a$ International School for Advanced Studies (SISSA/ISAS)\\
Via Beirut 2--4, 34014 Trieste, Italy, and INFN, Sezione di
Trieste\\
  $^b$ Theoretical Physics Department, Faculty of Science,
        University of Zagreb\\
        p.p. 331, HR-10002 Zagreb, Croatia\\

E-mail:   \email{bonora@sissa.it}, \email{mcvitan@phy.hr}, \email{pallua@phy.hr}, \email{ismolic@phy.hr}}

\abstract{We complete the analysis carried out in previous papers by studying
the Hawking radiation for a Kerr black--hole carried to infinity by fermionic 
currents of any spin. We find agreement with thermal spectrum of 
the Hawking radiation for fermionic degrees of freedom.
We start by showing that the near--horizon physics for a Kerr black--hole
is approximated by an effective two--dimensional field theory of fermionic
fields. Then, starting from 2d currents of any spin that form a
$W_{1+\infty}$ algebra, we construct an infinite set of covariant currents,
each of which carry the corresponding moment of the Hawking radiation.
All together they agree with the thermal spectrum of the latter.
We show that the predictive power of this method is not based on the anomalies
of the higher spin currents (which are trivial), but on the
underlying $W_{1+\infty}$ structure. Our results point toward
the existence in the near--horizon geometry of a symmetry larger than the Virasoro 
algebra, which very likely
takes the form of a $W_\infty$ algebra.}


\begin{document}

\maketitle

\section{Introduction}

This paper is complementary to \cite{BC} and \cite{BCPS}, referred to henceforth as
I and II, respectively. The subject of these previous papers was the
calculation of the Hawking radiation and its thermal spectrum by the method of anomalies
and the role played by a $W_\infty$ algebra of currents in this derivation.

Hawking radiation \cite{Hawking1,Hawking2}  does not depend on the details
of the collapse that gives rise to a black hole. Therefore one expects
that the methods to calculate it should have the same character of universality.
The anomaly method has these features. The first attempt
to compute Hawking radiation by exploiting trace anomalies was made
by Christensen and Fulling, \cite{CF} (see also \cite{Davies}), and reproposed subsequently
by \cite{Thorlacius,Strominger} in a modified form. More recently
a renewed attention to the same problem has been pioneered by the paper
\cite{Robinson}, which makes use of the diffeomorphisms anomaly.
This paper is at the origin of a considerable activity
with several contributions \cite{IUW1,IUW2,IMU1,IMU2,IMU3,
IMU4,IMU5,
Murata:2006pt,
Vagenas:2006qb,
Setare:2006hq,
Jiang:2007gc,
Jiang:2007pn,
Jiang:2007wj,
Kui:2007dy,
Shin:2007gz,
Jiang:2007mi,
Das:2007ru,
Chen:2007pp,
Miyamoto:2007ue,
Jiang:2007pe,
Kim:2007ge,
Murata:2007zr,
Peng:2007nj,
Ma:2007xr,
Huang:2007ed,
Peng:2008ru,
Wu:2008yx,
Gangopadhyay:2008zw,
Kim:2008hm,
Xu,Banerjee1,Banerjee2,Gango,Gango2,Kulkarni,Peng1,Peng2,Peng3,Iso:2008sq,Umetsu:2008cm,
Shirasaka:2008yg,Ghosh:2008tg,Banerjee:2008ez,Banerjee:2008az,
Akhmedova:2008au, Banerjee:2008fz, Gangopadhyay:2008ub, Banerjee:2008wq, Morita:2008qn,
Porfyriadis:2008yp, Banerjee:2008sn, Papantonopoulos:2008wp, Wei:2009kg, Nam:2009dd,
Fujikawa:2009bg, Morita:2009mt, Porfyriadis:2009zs, Wei:2009zt, Peng:2009uh}.

Most of these papers are concerned with the derivation of the integrated Hawking
radiation and do not describe its spectrum. However one of the most interesting features
of the Hawking radiation is precisely its thermal spectrum. The latter can be
`Fourier
analyzed' and expressed in terms of its higher moments or fluxes.
An interesting proposal was made by the authors of \cite{IMU1,IMU2,IMU3,IMU4},
who attributed these higher fluxes to phenomenological higher spin currents,
i.e. higher spin generalizations of the energy--momentum tensor.

In \cite{BC} it was shown that such higher currents do describe the higher spin
fluxes of the Hawking radiation. The main result of I was that this is not
due to their trace anomalies,
but rather the their underlying $W_\infty$ algebra structure. In fact it was
shown in \cite{BC} that
these higher spin currents cannot have trace anomalies and in \cite{BCPS}
that they cannot have diffeomorphism anomalies (or, rather, that if there
are anomalies they are trivial). In I and II the analysis was limited to
bosonic higher spin currents.
In the present paper we would like to extend the analysis to fermionic currents.
Our conclusion will not change: the thermal spectrum of the Hawking radiation
is induced not by the anomalies of such currents, which do not exist, but by
their underlying $W_{1+\infty}$ structure (the 1 stands for the extension of
the $W_\infty$ algebra to include a U(1) current). We will also examine some
aspects of the $W_{1+\infty}$ algebra which were not duly clarified in I and
II, but are basic to appreciate
the central role of the $W_\infty$ algebra. The main conclusion of our series of papers
is that the Hawking radiation and, in particular its thermal spectrum, points
toward the existence in the near horizon region of a symmetry much larger than
the Virasoro algebra, that is a $W_\infty$ or a $W_{1+\infty}$ algebra.

In this paper we will start (section 2) from a Kerr--like metric in 4D and consider
fermionic matter coupled to it and to a background electromagnetic field.
Like in I and II we will reduce the problem to two dimensions.
This can be done by using
azimuthal symmetry and the near horizon properties in the Kerr background. 
The spinor field, $\psi(t,r,\theta,\varphi)$ will be expanded
in the appropriate spherical harmonics. After integrating
the action over the polar angles one is left with infinite many free
two--dimensional spinor fields interacting with the background gravity
specified by the metric
\be
ds^2= f(r) dt^2 - \frac 1{f(r)}dr^2 \label{2metric}
\ee
as well as to the electromagnetic field. $f(r)$ near the horizon behaves
like $f(r)\approx 2 \k (r-r_H)$, where $\k$ is the surface gravity.
In the following we will focus on one of these complex fermion
fields. The analysis for all the other fermion fields is the same, what
is left out from our analysis is the resummation of all these contributions
and obtain the relevant four--dimensional information, see for instance
\cite{Fabbri}. After section 2 the
paper is organized as follows. In section 3 we recall
the trace anomaly method, which is basic in this paper. In section 4 
we introduce the currents of the $W_{1+\infty}$ algebra relevant to our problem. 
In section 5 we construct the covariant higher spin currents and show that
their flux at infinity is in agreement with the moments of the fermionic
Hawking radiation. In section 6 we discuss the problem of trace anomalies in higher
spin currents and, like in (\cite{BC,BCPS}), show on general grounds that there 
cannot be trace anomalies in these currents in accord with our explicit construction
in the previous section. Finally in section 7 we draw our conclusions.
Two Appendices are devoted to some details of the calculations in section 2 and 5.

\section{Reduction to two dimensions}

We start with the $4$-dimensional action for fermions in a curved background:
\be \label{faction}
S 
= \int d^4x \sqrt{-g} \bar\psi \displaystyle{\not} \nabla  \psi
= \int d^4x \sqrt{-g} \psi^\dagger \gamma^0 \gamma^a
e_a{}^\mu \left(  \partial_\mu -\frac{1}{8} \omega_{bc\mu} \left[ \gamma^b, \gamma^c \right] \right) \psi
\ee
where the vierbein $e^a{}_\mu{}$ satisfies $\eta_{ab} e^a{}_\mu{} e^b{}_\nu{} = g_{\mu\nu}$, and the spin 
connection $\omega^a{}_{b\mu}$ is given by $\omega^a{}_{b\mu} = e^a{}_\nu \nabla_\mu e_b{}^\nu$. 
(Indices $a,b,c=0,1,2,3$ are flat, indices $\mu,\nu = t,r,\theta,\phi$ are curved.)

We consider the Kerr metric, 
\be \label{4metric}
ds^2 &=& \frac{\Delta}{\Sigma} \left(dt - a \sin^2\theta d\phi\right)^2 
- \frac{\sin^2\theta}{\Sigma} \left( a dt - \left(r^2 + a^2\right) d\phi \right)^2
\CR
&& - \left(r^2 + a^2 \cos^2 \theta\right)\left( \frac{dr^2}{\Delta} + d\theta^2 \right)
\ee
and we choose the following local Lorentz frame (i.e.\ the vierbein) $e_a{}^\mu$:
\be \label{4bein}
\sqrt{\Delta \Sigma} \; e_0{}^\mu \partial_\mu &=& \left(r^2+a^2\right) \partial _t+a \partial _{\phi } \CR
\sqrt{\Delta \Sigma} \; e_1{}^\mu \partial_\mu &=&  \Delta  \partial_r \CR
\sqrt{\Delta \Sigma} \; e_2{}^\mu \partial_\mu &=&  \sqrt{\Delta } \partial _{\theta } \CR
\sqrt{\Delta \Sigma} \; e_3{}^\mu \partial_\mu &=&  \sqrt{\Delta } \left(a  \sin\theta  \partial_t 
+\frac{\partial_{\phi }}{\sin\theta } \right) 
\ee
where $\Sigma = r^{2} + a^{2}\cos^{2}\theta$, $\Delta = (r-r_{+})(r-r_{-})$, $r_++r_-=2M$, $r_+r_-=a^2$.
Near the horizon we have $r \rightarrow r_{+}$ and consequently $\Delta\rightarrow 0$. 
From the third and the fourth line of \refb{4bein} we see that the terms in the action 
\refb{faction} which are multiplied by $\gamma^2 e_2{}^\mu$ and $\gamma^3 e_3{}^\mu$ 
are suppressed by a factor of $\sqrt{\Delta}$. We can see that the term 
$\gamma^1 e_1{}^\mu \partial_\mu$ is not suppressed, by changing to tortoise 
coordinate $r^*$ defined by $\frac{dr^*}{dr} = \frac{r^2 + a^2}{\Delta}$. 
Expressed in terms of $r^{*}$, $\sqrt{\Delta \Sigma} \; e_1{}^\mu \partial_\mu$ becomes
$(r^2 + a^2)  \partial_{r^*}$.
Therefore, the leading order contribution from the term 
$\gamma^a e_a{}^\mu \partial_\mu$ in the action \refb{faction} is 
$\gamma^0 e_0{}^t \partial_t + \gamma^0 e_0{}^\phi \partial_\phi + 
\gamma^1 e_1{}^r \partial_r$, and is of order $1/\sqrt{\Delta}$.
Furthermore, a straightforward calculation shows that the leading contribution of 
the term $e_c{}^\mu \omega_a{}_b{}_\mu$ comes from
$e_0{}^\mu \omega_{01\mu} = -e_0{}^\mu \omega_{10\mu} = \frac{r_{+} - 
r_{-}}{2\sqrt{\Delta\Sigma}}$ and is also of order $1/\sqrt{\Delta}$ 
(the spin coefficients are listed in the Appendix~A).

In summary, on the horizon $r \approx r_+$, we obtain in the leading order
\be
\displaystyle{\not} \nabla \psi = \left\lbrace
\frac{\gamma^0}{\sqrt{\Delta \Sigma}} 
\left[ \left( r_{+}^2+a^2 \right) \partial_t + a \partial_\phi \right]  + 
\frac{\gamma^1}{\sqrt{\Delta \Sigma}} 
\left[ \left( r_{+}^2+a^2 \right) \partial_{r^*} - \frac{1}{4}\left( r_{+} - 
r_{-} \right) \right] \right\rbrace \psi 
\ee

To be able to integrate over $\theta$ and $\phi$ in the action \refb{faction}, we 
expand $\psi$ in the following way $\psi = \sum_{lm} \psi_{lm}(t,r) S_{lm}(\theta) 
e^{-i m \phi}$, where $S_{lm}$ are normalized so that 
$\int d\theta \sqrt{\Sigma} \sin\theta S^{*}_{lm}(\theta) S_{l'm}(\theta) = 
2 \delta_{l l'}$. That produces the change $\partial_\phi \rightarrow -i m$. 
We first integrate over $\phi$, and then over $\theta$ using normalization 
condition for $S_{lm}$ and obtain:
\be
S=4\pi \int dt dr \frac{ r_{+}^2+a^2 }{\sqrt{\Delta}}
\sum_{lm} \psi^\dagger_{lm}  \left\lbrace
\gamma^0\gamma^0
\left(  \partial_t - \frac{i a m}{ r_{+}^2+a^2 } \right)  + 
\gamma^0\gamma^1
\left(  \partial_{r^*} - \frac{r_{+} - r_{-}} {4( r_{+}^2+a^2 ) } \right)
\right\rbrace  \psi_{lm} \nonumber
\ee
We choose the following gamma matrices in $4D$,
\be
\gamma^0 = 
	\left(
\begin{array}{cccc}
	\nn{0} & \nn{0} & \nn{0} & \nn{1} \\
	\nn{0} & \nn{0} & \nn{1} & \nn{0} \\
	\nn{0} & \nn{1} & \nn{0} & \nn{0} \\
	\nn{1} & \nn{0} & \nn{0} & \nn{0}
\end{array}
\right)&,&
	\gamma^1 = \left(
\begin{array}{cccc}
	\nn{0} & \nn{0} & \nn{0} & \nn{-1} \\
	\nn{0} & \nn{0} & \nn{1} & \nn{0} \\
	\nn{0} & \nn{-1} & \nn{0} & \nn{0} \\
	\nn{1} & \nn{0} & \nn{0} & \nn{0}
\end{array}
\right),\0\\
	\gamma^2 = \left(
\begin{array}{cccc}
	\nn{0} & \nn{0} & \nn{-i} & \nn{0} \\
	\nn{0} & \nn{0} & \nn{0} & \nn{i} \\
	\nn{-i} & \nn{0} & \nn{0} & \nn{0} \\
	\nn{0} & \nn{i} & \nn{0} & \nn{0}
\end{array}
\right)&,&
	\gamma^3 = 
\left(
\begin{array}{cccc}
	\nn{0} & \nn{0} & \nn{1} & \nn{0} \\
	\nn{0} & \nn{0} & \nn{0} & \nn{1} \\
	\nn{-1} & \nn{0} & \nn{0} & \nn{0} \\
	\nn{0} & \nn{-1} & \nn{0} & \nn{0}
\end{array}
\right)\0
\ee
and the following gamma matrices in $2D$
\begin{equation}
\sigma^0 = 
	\left(
\begin{array}{cc}
	\nn{0} & \nn{1} \\
	\nn{1} & \nn{0} 
\end{array}
\right),
	\sigma^1 = \left(
\begin{array}{cc}
	\nn{0} & \nn{-1} \\
	\nn{1} & \nn{0} 
\end{array}
\right)
\end{equation}
The choice ensures that $\gamma^0 \gamma^1$ and 
$\sigma^0 \sigma^1$ look very simple. Both are diagonal, 
and satisfy $\gamma^0 \gamma^1 = I \otimes \sigma^0 \sigma^1$. 
Since $\gamma^0 \gamma^1$ and $\sigma^0 \sigma^1$ generate 
$01$-Lorentz transformation in $4D$ and $2D$ respectively, 
the upper two, as well as the lower two components of $\psi_{lm}$, will transform 
like the $2D$ spinors.
We denote the two upper components 
by $\chi_{(1)lm}$ and the two lower by $\chi_{(2)lm}$:
\be
\psi_{lm} = \left( \begin{array}{c} 
	\chi_{(1)lm} \\ \chi_{(2)lm} \end{array}\right)
\ee
In terms of $\chi_{(s)lm}$ ($s=1,2$) the action reads
\be
S &=& 4\pi \int dt dr\, \frac{ r_{+}^2+a^2 }{\sqrt{\Delta}}\0\\
&&\cdot \sum_{s=1}^2\sum_{lm} \chi^\dagger_{(s)lm}  \left\lbrace
\sigma^0\sigma^0
\left(  \partial_t - \frac{i a m}{ r_{+}^2+a^2 } \right)  + 
\sigma^0\sigma^1
\left(  \partial_{r^*} - \frac{r_{+} - r_{-}} {4( r_{+}^2+a^2 ) } \right)
\right\rbrace  \chi_{(s)lm} \nonumber
\ee

Now we show that we can interpret the action in terms of $2D$ quantities: 
the spinors $\chi_{(s)lm}$, the metric \refb{2metric}, its zweibein 
$e_i^{(2)}{}^\alpha$ and spin connection $\omega_{jk\alpha}^{(2)}$, a vector 
potential $A_\alpha$ and a dilaton $\Phi$. We take the letters $i,j,k=0,1$ to 
denote flat $2D$ indices, and $\alpha=t,r$ to denote the curved. First we calculate 
the $2D$ covariant derivative ${}^{(2)}\nabla_\alpha$ contracted with $2D$ gamma 
matrices $\sigma^i e_i^{(2)\alpha}$
\be \label{2nablaslash}
{}^{(2)}{\displaystyle{\not} \nabla} \chi= 
\sigma^i e_i^{(2)}{}^\alpha \left(  \partial_\alpha  - \frac{1}{8} 
\omega_{jk\alpha}^{(2)}\left[ \sigma^j, \sigma^k \right] \right) \chi
 = \left\{ 
\frac{ \sigma^0}{\sqrt{f(r)}} \partial_t +
\frac{ \sigma^1}{\sqrt{f(r)}} 
\left[ \partial_{r^*} - \frac{f'(r)}{4} \right]  \right\}\chi
\ee
Next, motivated by the fact that for the $4D$ metric \refb{4metric} the tortoise 
coordinate satisfies $\frac{dr^*}{dr} = \frac{r^2 + a^2}{\Delta}$, whereas for 
the $2D$ metric \refb{2metric} it satisfies $\frac{dr^*}{dr} = \frac{1}{f(r)}$, 
we identify
\be
f(r) = \frac{\Delta(r)}{ r^2+a^2 }
\ee
Finally, plugging this into \refb{2nablaslash}, we see that in the leading order 
near the horizon we can write the action in the following way
\be
S = \sum_{s=1}^2\sum_{lm} 4\pi \int dt dr \, \Phi
 \bar\chi_{(s)lm}  
{\displaystyle{\not} D}
\chi_{(s)lm}
\ee
where the covariant derivative now includes the gauge part $D_\alpha = 
{}^{(2)}\nabla_\alpha - i q A_\alpha$, and the charge $q$ of $\chi_{(s)lm}$ is $m$. 
This is the $2D$ action for an infinite number of two component fermions $\chi_{(s)lm}$ 
in the background given by the dilaton $\Phi$
\be
\Phi = \sqrt{ r^2+a^2 }
\ee
the gauge field $A_\alpha$
\be
A_{t} &=&  \frac{e_0{}^\phi}{e_0{}^t} = \frac{a}{ r^2+a^2 }\CR
A_{r} &=& 0
\ee
and the metric \refb{2metric}.

In the sequel we restrict our analysis to the near horizon region. In this region the
dilaton is approximately constant, so we may disregard it: the equations of motion are
those of free fermions in two dimensions, coupled to the metric and the gauge field
(but not to the dilaton).

\section{The trace anomaly method}

To start with let us recall the trace anomaly method to compute the integrated
Hawking radiation (in the absence of a gauge field). With reference to the metric 
\refb{2metric} we transform it into a conformal metric by means of the 'tortoise'
coordinate $r_*$ defined via $\frac {\partial r}{\partial r_*}=f(r)$. Next
we introduce light--cone coordinates $u=t-r_*, v=t+r_*$. Let us denote
by $T_{uu}(u,v)$ and $T_{vv}(u,v)$ the classically non vanishing
components of the energy--momentum tensor in these new coordinates.
Our black hole problem is now reduced to the background metric
$g_{\alpha\beta} = e^\varphi \eta_{\alpha\beta}$, where $\varphi= \log f$.
The energy--momentum tensor can be calculated by integrating the conservation
equation and using the trace anomaly. The result is (see II)
\be
T_{uu}(u,v)= \frac {\hbar c_R} {24\pi} \left(\d_u^2 \varphi-
\frac 12 (\d_u\varphi)^2\right)+T^{(hol)}_{uu} (u)\label{TuuThol}
\ee
where $T_{uu}^{(hol)}$ is holomorphic, while $T_{uu}$ is conformally
covariant. Namely, under a conformal transformation $u\to \tilde u=f(u)
(v\to \tilde v=g(v))$ one has
\be
T_{uu}(u,v)=\left(\frac {df}{du}\right)^2 T_{\tilde u\tilde u}
(\tilde u, v)\label{conf}
\ee
Since, under a conformal transformation, $\tilde \varphi(\u,\tilde v)=
\varphi(u,v) -\ln \left( \frac {df}{du}\frac {dg}{dv}\right)$, it follows
that
\be
T^{(hol)}_{\u \u}(\u) = \left(\frac {df}{du} \right)^{-2} \left(
T^{(hol)}_{u u}(u)+ \frac {\hbar c_R}{24\pi} \{\u,u\}\right)\label{Thol}
\ee
Let us pass to Kruskal coordinates, which are regular at the horizon,
i.e. to $(U,V)$ defined by
$U=-e^{-\k u}$ and $V=e^{\k v}$. Under this transformation we have
\be
T^{(hol)}_{UU}(U) = \left(\frac 1{\k U} \right)^{2} \left(
T^{(hol)}_{u u}(u)+ \frac {\hbar c_R}{24\pi} \{U,u\}\right)\label{TholU}
\ee
Now we require the outgoing energy flux to be regular at the future
horizon $U=0$ in the Kruskal coordinate. Therefore at that point
$T^{(hol)}_{u u}(u)$ is given by $\frac  {c_R \k^2}{48 \pi}$. As was noticed
in \cite{BCPS} this requirement corresponds to the condition that $T_{uu}(u,v)$
vanishes at the horizon.

Since the background is static, $T^{(hol)}_{u u}(u)$
is constant in $t$ and therefore also in $r$. Therefore at $r=\infty$
it takes the same value $\frac {\hbar c_R \k^2}{48 \pi} $.
On the other hand we can assume that at $r=\infty$
there is no incoming flux and that the background is
trivial  (so that the vev of $T^{(hol)}_{u u}(u)$ and $T_{u u}(u,v)$
asymptotically coincide).

Therefore the asymptotic flux is (we denote by $\langle \cdot\rangle$ the
value at infinity)
\be
\langle T_t^r\rangle =\VEV {T_{uu}}-\VEV {T_{vv}} =
\frac{\hbar \k^2}{48 \pi} c_R\label{flux}
\ee
This is the integrated Hawking radiation (see below).

We would like to apply a similar method to the higher spin currents.
Let us start by recalling a few notions about the thermal fermionic radiation.

The thermal fermionic spectrum of the Kerr black hole is given by the Planck distribution
$$N(\omega) = \frac{1}{e^{\beta (\omega-m \Omega)} + 1}$$
where $1/\beta$ is Hawking temperature of the black hole, $\omega$ is the absolute
value of the momentum ($\omega = |k|$) and $\Omega$ is the total angular momentum,
in our case $\Omega= A_t$ evaluated at the horizon and $m$ is the charge.

Let us consider first the case $m=0$. In two dimensions we can define the flux moments
$F_n$, which vanish for $n$ odd, while for $n$ even they are given by, \cite{IMU5},
\be
F_{2n} = \frac{1}{2\pi} \int_{0}^{\infty} d\omega \,
\frac{\omega^{2n-1}}{e^{\beta \omega} + 1} =  \frac{\kappa^{2n} B_{2n}}{8\pi n} \,
 (1 - 2^{1-2n}) (-1)^{n+1}\label{F2nf}
\ee
where $B_s$'s are the Bernoulli numbers $(B_2 = 1/6 \ , \ B_4 = -1/30 \ , \ \dots)$
and $\kappa = 2\pi/\beta$ is the surface gravity of the black hole.

When $m\neq 0$ we do not have similar compact formulas, however it makes sense to sum
over the emission of a particle (with charge $m$) and the corresponding antiparticle 
(with charge $-m$). In this case the flux moments become
\be
F^{\Omega}_{n+1}&=&\frac 1{2\pi}\left(\int_0^\infty dx \,
\frac{x^n}{e^{\beta (x-m\Omega)}+1}-(-1)^n 
\int_0^\infty dx \,\frac{x^n}{e^{\beta (x+m\Omega)}+1}\right)\0\\
&=& \frac {(m\Omega)^{n+1}}{2\pi(n+1)} -\sum_{k=1}^{[(n+1)/2]} (-1)^k
\frac {n!\,(1-2^{1-2k})\kappa^{2k}}{2\pi (2k)!(n+1-2k)!} B_{2k} (m\Omega)^{n+1-2k}\label{Fn}
\ee
Once we know $F_n^\Omega$ we do not have enough information to reconstruct the full 
thermal spectrum with $m\neq 0$, but being able to reproduce the moments $F_n^\Omega$ 
represents anyhow an important positive test.

\section{A $W_{1+\infty}$ algebra}

In order to derive the higher Hawking fluxes the same way we derived above the
integrated Hawking radiation, we postulate the existence of conserved spin currents
consisting of fermionic bilinears in the 2D effective field theory near the horizon.
They will play a role analogous to the energy--momentum tensor for the integrated
radiation (the lowest moment). To construct such currents we start from
a $W_{1+\infty}$ algebra defined in an abstract flat space spanned by a local
coordinate $z$. These currents were introduced in \cite{Pope3} (see also \cite{BK,Pope1,Pope2}):
\be
j_{z\ldots z}^{(s)} (z) &=& - \frac{B(s)}{s} \, \sum_{k=1}^{s} (-1)^k {s-1 \choose s-k}^{2} :
 \partial_z^{s-k} \Psi^{\dagger}(z) \, \partial_z^{k-1} \Psi(z) :\label{js}\\
&&B(s) \equiv \frac{2^{s-3} s!}{(2s-3)!!} \, q^{s-2} \qqd s = 1, 2, 3, \dots \label{Bs}
\ee
where $q$ is a deformation parameter.

The spin $s$ currents $j^{(s)}_{z \dots z} (z)$ are linear combinations of bilinears
$$j^{(m,n)}_{z \dots z}(z) \equiv \ : \partial^m \Psi^{\dagger} \, \partial^n \Psi : \
\equiv \lim_{z_1,z_2\to z} \left( \partial_{z_1}^m \Psi^{\dagger}(z_1) \,
 \partial_{z_2}^n \Psi(z_2) - \partial_{z_1}^m \partial_{z_2}^n
\left< \Psi^{\dagger}(z_1) \Psi(z_2) \right> \right)$$
We want to relate the currents written in two different coordinate systems,
connected by coordinate change $z \rightarrow w(z)$. 
That is, we would like to obtain a relation analogous to the one found in \cite{BC} 
\begin{equation}\label{eq2}
j^{(s)}_{z\ldots z}(z) \rightarrow  
\left( \frac 1{\kappa w}\right)^s  \left(j^{(s)}_{z\ldots z}
+\VEV{ X^F_s}\right) 
\end{equation}
and apply it to the transformation $w(z)= -e^{-\k z}$ so as to obtain the value of
$j^{(s)}_{z\ldots z}(z)$ at the horizon by requiring regularity.

The following transformation
property of holomorphic fermionic fields will be needed
$$\Psi(z) = (w'(z))^{1/2} \, \Psi(w)$$ 

Using it we get
\be
&&:\partial^{m}_{z_1} \Psi^{\dagger}(z_1) \, \partial^{n}_{z_2} \Psi(z_2): \ =
\partial^{m}_{z_1} \, \partial^{n}_{z_2} : \Psi^{\dagger}(z_1) \, \Psi(z_2): \0\\
 &=&
\partial^{m}_{z_1} \, \partial^{n}_{z_2} \left( \Psi^{\dagger}(z_1) \, \Psi(z_2)
 - \left< \Psi^{\dagger}(z_1) \, \Psi(z_2) \right> \right) \0\\
&=& \partial^{m}_{z_1} \, \partial^{n}_{z_2} \left( (w'_{1}(z_1))^{1/2}
(w'_{2}(z_2))^{1/2} \Psi^{\dagger}(w_1) \, \Psi(w_2) - \left< \Psi^{\dagger}(z_1) \,
 \Psi(z_2) \right> \right) \0\\
&=& \partial^{m}_{z_1} \, \partial^{n}_{z_2} \left( (w'_{1}(z_1))^{1/2}
(w'_{2}(z_2))^{1/2} : \Psi^{\dagger}(w_1) \, \Psi(w_2) : \right) \0\\
&&+ \partial^{m}_{z_1} \, \partial^{n}_{z_2} \left( (w'_{1}(z_1))^{1/2}
(w'_{2}(z_2))^{1/2} \left< \Psi^{\dagger}(w_1) \, \Psi(w_2) \right> -
\left< \Psi^{\dagger}(z_1) \, \Psi(z_2) \right> \right)\0
\ee
Let us set
\be
G(z_1, z_2) \equiv - \left( (w'_{1}(z_1))^{1/2} (w'_{2}(z_2))^{1/2}
\left< \Psi^{\dagger}(w_1) \, \Psi(w_2) \right> - \left< \Psi^{\dagger}(z_1) \,
\Psi(z_2) \right> \right)\label{Gz1}
\ee
Then
\be
&&:\partial^{m}_{z_1} \Psi^{\dagger}(z_1) \, \partial^{n}_{z_2} \Psi(z_2): =\ \label{Gz2}\\
\quad&=&
\partial^{m}_{z_1} \, \partial^{n}_{z_2} \left( (w'_{1}(z_1))^{1/2} (w'_{2}(z_2))^{1/2}
: \Psi^{\dagger}(w_1) \, \Psi(w_2) : \right) - \partial^{m}_{z_1} \,
 \partial^{n}_{z_2} G(z_1, z_2)
\ee

The propagator for fermionic holomorphic fields is given by
\be
\left< \Psi^{\dagger}(z) \Psi(w) \right> = \frac{\lambda}{z - w}
\label{prop}
\ee
The value of $\lambda$ is determined in such a way as to reproduce
the transformation properties of the energy--momentum tensor and,
in physical units, is proportional to $\hbar$. Eventually we will set $\lambda=\hbar$.
We are interested in the transformation properties of fermionic currents when $w(z)$ is
$w(z) = - e^{-\kappa z}$.
Using this we have
\be
G(z_1, z_2) = G(z_1 - z_2) =
-\lambda \left( \frac{\kappa/2}{\sinh\left(\frac{\kappa}{2}(z_1 - z_2)\right)}
- \frac{1}{z_1 - z_2} \right)\label{Gz3}
\ee
Proceeding with our currents (\ref{js}) we obtain
\be
j_{z\ldots z}^{(s)}(z) &=& - \frac{B(s)}{s} \, \sum_{k=1}^{s} (-1)^k {s-1 \choose s-k}^{2}\0\\
&&\cdot\lim_{z_1 \rightarrow z_2} \partial^{s-k}_{z_1} \partial^{k-1}_{z_2}
\left( (w'_{1}(z_1))^{1/2} (w'_{2}(z_2))^{1/2} : \Psi^{\dagger}(w_1) \,
\Psi(w_2) : \right) + \left< X^F_s \right>\label{currents}
\ee
where
\be
\left< X^{F}_{s} \right> \equiv - \frac{B(s)}{s} \,
\sum_{k=1}^{s} (-1)^{k+1} {s-1 \choose s-k}^{2} \lim_{z_1 \rightarrow z_2}
\partial^{s-k}_{z_1} \partial^{k-1}_{z_2} G(z_1, z_2)\label{XF}
\ee
Now, using the familiar series
\be
\frac{a}{\sinh(ax)} - \frac{1}{x}  = -
 \sum_{p=1}^{\infty} \frac{a^{2p} (2^{2p-1} - 1) B_{2p}}{p} \,
 \frac{x^{2p-1}}{(2p-1)!}\0
\ee
for $a = \kappa/2$ we obtain
\be
\left< X^{F}_{s} \right> = - \frac{B(s)}{s} \,
\sum_{k=1}^{s} (-1)^{k+1} {s-1 \choose s-k}^{2} (-1)^{k-1}
 \lambda \frac{\kappa^{s} (1 - 2^{-(s-1)}) B_s}{s}\label{XF1}
\ee
Finally, using the value of the sum,
$$\sum_{k=1}^{s} {s-1 \choose s-k}^{2} = 2^{s-1} \frac{(2s-3)!!}{(s-1)!}$$
we find
\be
\left< X^{F}_{s} \right> = -\lambda \frac{\kappa^s B_s}{s} \,
(1 - 2^{1-s}) (4q)^{s-2}= - \VEV{j^{(s)}_{z\ldots z}}_h\label{XFs}
\ee
where $\VEV{\cdot}_h$ denotes the value at the horizon.
Notice that $\left< X^F_s \right> = 0$ for an odd spin $s$. For $s>1$ this is because 
$B_{s} = 0$ for odd $s>1$. For $s=1$ it is because of the other factor in \refb{XFs}.

\vspace{1.0cm}

\section{Higher spin covariant currents}

The holomorphic currents of the previous section refer to a background with
a trivial Euclidean metric. In order to construct the corresponding
covariant higher-spin currents from fermionic fields, first
we recall some properties of fermions in two dimensions \cite{IMU4}.
The equation of motion for
a right-handed fermion with unit charge is given by
\be
\left( \partial_u - iA_v + \frac{1}{4}\,\partial_v \varphi \right) \psi(u,v) = 0
\label{eom}
\ee
In the Lorentz gauge, the gauge field can be written locally as
$A_u = \partial_u \eta(u,v)$ and $ A_v = -\partial_v \eta(u,v)$
where $\eta(u,v)$ is a scalar field. Since gravitational and gauge fields are not
generally holomorphic, $\psi(u,v)$ is not holomorphic either. In order to construct
holomorphic quantities from a fermionic field, we define a new field $\Psi$ by
\be
\Psi \equiv \exp\left(\frac{1}{4}\,\varphi(u,v) + i\eta(u,v) \right) \psi(u,v)\label{Psi}
\ee
It is easy to show that the equation of motion implies $\partial_v \Psi = 0$ and hence
$\Psi$ is holomorphic. Similarly we can define $\Psi^{\dagger}$ as
\be
\Psi^{\dagger} \equiv \exp\left(\frac{1}{4}\,\varphi(u,v) - i\eta(u,v) \right)
\psi^{\dagger}(u,v)\0
\ee
The equation of motion again guarantees that $\partial_v \Psi^{\dagger} = 0$, so that
$\Psi^{\dagger}$ is also holomorphic. We will use $\Psi$ and $\Psi^\dagger$
as the basic chiral fields to construct the $W_{1+\infty}$ algebra introduced in the
previous section. To covariantize the expressions of the currents we reduce the
problem to one dimension by considering only the $u$ dependence and keeping
$v$ fixed. In one dimension a curved coordinate $u$ in the presence of a background
metric
$$g_{\mu \nu} = e^{\varphi(u,v)} \eta_{\mu \nu}$$
is easily related to the
corresponding normal coordinate $x$ by the equation
$\partial_x = e^{-\varphi(u,v)} \partial_{u}$. We view $u$ as $u(x)$
and, by the
above equation, we extract the correspondence between $j^{(s)}_{z\ldots z}$
and $j^{(s)}_{u\ldots u}$ by identifying $u$ with 
the coordinate $z$ of the previous section after Wick rotation. 
The expressions we get in this way are not yet
components of the covariant currents.
We have to remember the current conformal weights and
introduce suitable factors in order to take them into account.

Under a holomorphic conformal transformation $u \rightarrow \tilde{u}$ the function
$\varphi(u,v)$ and the field $\Psi(u)$ transform according to
\be
\tilde{\varphi}(\tilde{u},v) &=& \varphi(u,v) - \ln\left(\frac{d\tilde{u}}{du}\right)\0\\
\tilde{\Psi}(\tilde{u}) &=& \left(\frac{d\tilde{u}}{du}\right)^{\frac{1}{2}} \Psi(u)\0
\ee
Therefore $e^{-\varphi/2} \Psi(u)$ (and analogously, $e^{-\varphi/2}
\Psi^{\dagger}(u)$) transforms as a scalar with respect to a holomorphic coordinate
transformation.

A remark is in order about the transformation property of the fermion field
$\Psi$ under (holomorphic) gauge transformations; in the Lorentz gauge there remains a
residual holomorphic gauge symmetry,
\be
\psi'(u,v) = e^{i\Lambda(u)} \psi(u,v) \qqd \eta'(u,v) = \eta(u,v) + \Lambda(u)\0
\ee
Under this transformation the field $\Psi(u)$ transforms as a field with twice the charge
of $\psi$, i.e. $\Psi'(u) = e^{2i\Lambda(u)} \Psi(u)$.

As a consequence the covariant derivative of $\Psi(u)$ turns out to be
\be
\nabla_u \Psi(u) &=& \left( \partial_u - \frac{1}{2} \partial_u \varphi -
2i A_u \right) \Psi(u)\0\\
\nabla_u \Psi^{\dagger}(u) &=& \left( \partial_u - \frac{1}{2} \partial_u \varphi +
2i A_u \right) \Psi^{\dagger}(u)\0
\ee
and for higher covariant derivatives we have,
\be
\nabla^{m+1}_u \Psi(u) &=& \left( \partial_u - \left( m +
\frac{1}{2} \right) \partial_u \varphi - 2i A_u \right) \nabla^{m}_u \Psi(u)\label{nablapsi}\\
\nabla^{m+1}_u \Psi^{\dagger}(u) &=& \left( \partial_u - \left( m + \frac{1}{2} \right)
 \partial_u \varphi + 2i A_u \right) \nabla^{m}_u \Psi^{\dagger}(u)\label{nablapsidag}
\ee
It can be shown that $e^{-(m+\frac{1}{2})\varphi} \, \nabla_u^{m} \Psi(u)$ and
$e^{-(m+\frac{1}{2})\varphi} \, \nabla_u^{m} \Psi^{\dagger}(u)$ transform as scalars
under holomorphic coordinate transformation, for every $m \in \mathbb{N}$.

After these preliminaries the covariant currents are constructed using the following
bricks:
\be
J_{u \dots u}^{(m,n)} &=& e^{(m+n+1)\varphi(u,v)} \lim_{\epsilon \rightarrow 0}
 \left( 
\vphantom{\frac{c_{m,n}^{f}}{\epsilon^{m+n+1}}}
e^{2i \int_{u_\mathrm{-}}^{u_\mathrm{+}} \! A_u (u',v) du'} \right.\label{Juumn}\\
&&\cdot\left. e^{-(m+1/2)\varphi(u_{+},v)} \nabla_{u}^{m} \Psi^{\dagger}(u_{+}) \,
e^{-(n+1/2)\varphi(u_{+},v)} \nabla_{u}^{n} \Psi(u_{-}) -
 \frac{c_{m,n}^{f}}{\epsilon^{m+n+1}} \right)
\ee
where we have used the abbreviations $u_{+} \equiv u(x + \epsilon/2)$
and  $u_{-} \equiv u(x - \epsilon/2)$.
The numerical constants $c_{m,n}^{f}$, defined by
$$c_{m,n}^{f} = \lambda (-1)^m (m+n)!$$
are determined in such a way that all singularities are canceled in the final
expressions for $J^{(m,n)}$.

Finally, let us define the covariant currents corresponding
to the $W_{1+\infty}$ fermionic currents:
\be
J_{u\ldots u}^{(s)} &=& - \frac{B(s)}{s} \, \sum_{k=1}^{s} (-1)^k {s-1 \choose s-k}^{2}
J_{u \dots u}^{(s-k,k-1)}\label{Jsuu}\\
B(s) &\equiv& \frac{2^{s-3} s!}{(2s-3)!!} \, q^{s-2}\0
\ee

\vspace{0.5cm}

The first few covariant $W_{1+\infty}$ fermionic currents can be written in pretty
simple form, using the abbreviation $T \equiv \partial_{u}^{2}\varphi - \frac{1}{2} \,
(\partial_{u} \varphi)^{2}$
\be
J^{(1)}_{u} &=& j^{(1)}_{u} + \frac{i\lambda}{2q} A_{u}\label{J1}\\
J^{(2)}_{uu} &=& \left(2 A_u^2-\frac{T}{12}\right) \la -
2 A_u J^{(1)}_{u}+j^{(2)}_{uu}\label{J2}\\
J^{(3)}_{uuu} &=& -4 J^{(1)}_{u} A_u^2-4 J^{(2)}_{uu} A_u+\left(\frac{8 A_u^3}{3}-
\frac{A_u T}{3}\right) \la +\frac{T J^{(1)}_{u}}{6}+j^{(3)}_{uuu}\label{J3}\\
J^{(4)}_{uuuu} &=&
+\la  \left(4 A_u^4-\frac{7 T A_u^2}{5}-\frac{2}{5} \left(\nabla _u^2A_u\right) A_u+
\frac{7 T^2}{240}+\frac{3}{5} \left(\nabla _uA_u\right)^2\right)
\label{J4}\CR
&&-8 J^{(1)}_{u} A_u^3-12 J^{(2)}_{uu} A_u^2+\left(\frac{1}{5} \nabla _u^2J^{(1)}_{u}+
\frac{7 T J^{(1)}_{u}}{5}-6 J^{(3)}_{uuu}\right) A_u-\frac{3}{5} 
\left(\nabla _uA_u\right) \left(\nabla _uJ^{(1)}_{u}\right)\CR
&&+\frac{1}{5} \left(\nabla _u^2A_u\right) J^{(1)}_{u}+\frac{7 T J^{(2)}_{uu}}{10}+
j^{(4)}_{uuuu}
\ee
The expression for the fifth order current can be found in Appendix B. For the other 
currents we have explored up to order 8, the expressions are so unwieldy that we
have decided not to write them down explicitly.

\vspace{0.5cm}

Next we write down the covariant derivatives of the $W_{1+\infty}$ fermionic currents,
$J^{(s)}$, defined above,

\be
\label{nablaJ1}
g^{uv}\nabla_vJ^{(1)}_{u} &=& -\la F_u{}^u  \\
g^{uv}\nabla_vJ^{(2)}_{uu} &=& \frac{1}{24} \la
\left(\nabla_uR\right)+F_u{}^u J^{(1)}_{u}\label{nablaJ2}\\
g^{uv}\nabla_vJ^{(3)}_{uuu} &=& 2 F_u{}^u J^{(2)}_{uu}-\frac{1}{12}
\left(\nabla_uR\right) J^{(1)}_{u}\label{nablaJ3}\CR
g^{uv}\nabla_vJ^{(4)}_{uuuu} &=& \frac{3}{10} \left(\nabla_uF_u{}^u\right)
\left(\nabla_uJ^{(1)}_{u}\right)-
\frac{1}{10} F_u{}^u \left(\nabla_u^2J^{(1)}_{u}\right)-
\frac{1}{10} \left(\nabla_u^2F_u{}^u\right) J^{(1)}_{u}
\\
&&-\frac{7}{20} \left(\nabla_uR\right) J^{(2)}_{uu}+3 F_u{}^u J^{(3)}_{uuu}\label{nablaJ4}
\ee

In the case of lowest spin current, $J^{(1)}$, (\ref{nablaJ1}) gives rise to the
gauge anomaly
\be
g^{\mu\nu} \nabla_{\mu} J_{\nu}^{(1)} =
-\frac{\hbar}{2} \,
\epsilon^{\mu \nu} F_{\mu \nu}\label{gaugeanom}
\ee

Apart from the gauge anomaly in the first current we are interested to check whether
there are trace anomalies in the other currents. This is done as follows.
After the RHS of the above equation is expressed in terms of covariant quantities,
terms proportional to $\hbar$ (which is present only in $\lambda$)
are identified as possible anomalies by proceeding in analogy to the
energy--momentum tensor. One assumes that
there is no anomaly in the conservation laws of
covariant currents, that is that the covariant derivatives of the higher
spin currents with the addition of suitable covariant terms (these terms are
classical i.e. not proportional to $\hbar$, see for instance the terms in the LHS of
(\ref{nablaJ2},\ref{nablaJ3}) vanish. Since
$$\nabla \cdot J^{(m,n)}+\ldots = g^{uv} \nabla_{\!v} J_{u \dots u} + g^{uv}
\nabla_{\!u} J_{vu \dots u} +\ldots = 0,$$ where dots denote the
above mentioned classical covariant terms,
one relates terms proportional to $\hbar$ in the $u$ derivative of the trace
($_{vu \dots u}$ components) with the terms proportional to $\hbar$ in the $v$
derivative of $_{u \dots u}$ components of the currents.

For the covariant energy momentum tensor, $J^{(2)}$ we have
$\textrm{Tr}(J^{(2)}) = 2 g^{vu} J_{vu}^{(2)} = -\frac{\hbar}{12} R$
which is the well known trace anomaly. In the case of $J^{(3)}$ current the terms that
carry explicit factors of $\hbar$ cancel out in $g^{uv} \nabla_{\!v} J_{uuu}^{(3)}$,
which implies absence of $\hbar$ in the trace, and consequently the absence of the
trace anomaly. The same is true for $J^{(4)}$ and the higher currents.

\subsection{Higher moments of the Hawking radiation}

Now let us come to the description of the higher moments of the
fermionic Hawking radiation. We will follow the pattern outlined in section 3
and consider first the case in which the electromagnetic field is decoupled
($m=0$). 

In section 4 we evaluated $\VEV {j^{(s)}_{z\ldots z}}_h$.
If we identify $j^{(s)}_{z\ldots z}(z)$ via a Wick rotation 
with $j^{(s)}_{u\ldots u}(u)$ we get the corresponding
value at the horizon $\VEV {j^{(s)}_{u\ldots u}}_h$. We notice that since
the problem we are considering is stationary and
 $j^{(s)}_{u\ldots u}(u)$ is chiral, it follows that it is constant in $t$ and $r$. 
Therefore $\VEV {j^{(s)}_{u\ldots u}}_h$ corresponds to its value at $r=\infty$. 
Since $j^{(s)}_{u\ldots u}(u)$
and  $J^{(s)}_{u\ldots u}(u)$ asymptotically coincide, the asymptotic
flux of these currents is
\be
 \VEV{J^{(s)^r}{}_{t\ldots t}}=
  \VEV{J^{(s)}_{u\ldots u}}-
\VEV{J^{(s)}_{v\ldots v}}=  
\VEV {j^{(s)}_{u\ldots u}}_h \0
\ee
If we set $q=\frac i4$ and $\lambda=1$ in conventional units and, as in 
\cite{BC,BCPS} we multiply the currents by $-\frac 1{2\pi}$ in order to 
properly normalize the (physical) energy--momentum tensor, we get
\be
-\frac 1{2\pi} \VEV{J^{(2n)^r}{}_{t\ldots t}}=
-(-1)^n \frac{\kappa^{2n} B_{2n}}{4 \pi n} \,
(1 - 2^{1-2n})  \label{F2nflux}
\ee
while the odd currents give a vanishing value. These values correspond precisely
to the fluxes of the Hawking thermal spectrum defined by (\ref{F2nf}) 
multiplied by two. This is so because our currents carry both particle and antiparticle 
contributions.

Next we wish to take 
into account the presence of the gauge field, which, in our case, vanishes
at infinity but not at the horizon. This introduces a significant change in our 
method. In section 3 the basic criterion was to regularity of $T^{(hol)}_{uu}$ at
the horizon. Now the presence of the electromagnetic field interferes with the
regularity of $T^{(hol)}_{uu}$ at the horizon. As a consequence we have to update
our criterion.

Let us start with the first current (\ref{J1}). From now on we understand that
the electromagnetic field $A_u$ absorbs also the charge $m$, so that in the final 
results the replacement $A_t\to m\, A_t$ is understood. We easily get (remember that
$\langle X_1^F\rangle$ vanishes)
\be
J^{(1)}_{\u}= j_{\u}^{(1)} + \frac{i\lambda}{2q} A_{\u}=
\frac 1{f_u} \left(j_u^{(1)} + \frac{i\lambda}{2q} A_u\right)\label{Ju1}
\ee
where $f_u$ denotes the first derivative of $\u=f(u)$ with respect to $u$.
Now let us introduce the Kruskal coordinate $f(u)\equiv U=-e^{-\kappa u}$.
It is evident that we have to require regularity at the horizon of 
$j_\u^{(1)} + \frac{i\lambda}{2q} A_\u$, not of $j^{(1)}_\u$ alone. 
Therefore we get
\be
\VEV{j_\u^{(1)}}_h + \frac{i\lambda}{2q} \VEV{A_\u}_h=0\label{j1hor}
\ee
where $\VEV{\cdot}_h$ denotes the value at the horizon. Now $j_u^{(1)}(u)$ is constant
in $t$ and $r$. Therefore  
$\VEV{j_u^{(1)}}_h =- \frac{i\lambda}{2q} \VEV{A_u}_h$ corresponds to its value 
at $r=\infty$. Since $j^{(1)}_{u}(u)$
and  $J^{(1)}_{u}(u)$ asymptotically coincide, because $A_u(u)$ asymptotically vanishes,
we get
\be
-\frac 1{2\pi} \VEV{J^{(1)r}}=
-\frac 1{2\pi} \VEV{J^{(1)}_{u}}+
\frac 1{2\pi} \VEV{J^{(1)}_{v}}= \frac{i\lambda}{4\pi q}\VEV{A_u}_h= \frac 1{2\pi} A_t=
\frac {\Omega}{2\pi}\label{j1flux}
\ee
where $\VEV{\cdot}$ represents the asymptotic value and we have assumed that
there is no incoming flux $ \VEV{J^{(1)}_{v}}$ from infinity.

From this example we learn the obvious lesson.  
We have to assume that the currents $J^{(s)}_{U\ldots U}$ are regular on the horizon 
in Kruskal coordinates $U=-e^{-\kappa u}$.
Since these currents are covariant, we have 
$$J^{(s)}_{U\ldots U} =\frac 1{(-\kappa U)^s}\, J^{(s)}_{u\ldots u} (u)$$
It then follows that the currents $J^{(s)}_{u\ldots u}$, and 
their $n-1$ derivatives vanish. From \refb{J2}-\refb{J4}, at the horizon we must get
\be
j^{(2)}_{uu} &=& -\lambda \left(2 A_u^2-\frac{T}{12}\right) \0 \\
j^{(3)}_{uuu} &=& -\la\left(\frac{8 A_u^3}{3}-\frac{A_u T}{3}\right) \label{M2}\\
j^{(4)}_{uuuu} &=& -\la  \left(4 A_u^4-\frac{7 T A_u^2}{5}-\frac{2}{5} \left(\nabla _u^2A_u\right) A_u+
\frac{7 T^2}{240}+\frac{3}{5} \left(\nabla _uA_u\right)^2\right)\0
\ee
As already remarked, at infinity the background fields $A_u$ and $\phi$ vanish. So that 
\be
\VEV{J^{(s)}_{u\ldots u}} = \VEV{j^{(s)}_{u\ldots u}}_h \label{M3}
\ee
Now, we evaluate the derivatives on right hand side of \refb{M2} at the horizon.
Setting $\la=\hbar=1$ we get 
\be
 \VEV{j^{(2)}_{uu}}_h &=&  \frac{ \VEV{T}_h }{12}-\frac{ \VEV{A_t}_h^2}{2} \0\\
\VEV{j^{(3)}_{uuu}}_h &=& -\frac{1}{3}  \VEV{A_t}_h^3-\frac{1}{6} \VEV{T}_h  
\VEV{A_t}_h \0\\
\VEV{j^{(4)}_{uuuu}}_h &=& -\frac 14  \VEV{A_t}_h^4+\frac 14  \VEV{T}_h  
\VEV{A_t}_h^2-\frac{7}{240}  \VEV{T}_h ^2 \0
\ee
Therefore at infinity we get 
\be
-\frac{1}{2\pi} \VEV{J^{(2)r}_t} &=&  \frac{ \kappa ^2}{48 \pi }
+\frac{ \Omega ^2}{4 \pi } \0\\
-\frac{1}{2\pi} \VEV{J^{(3)r}_{tt}} &=&  \frac{ \Omega ^3}{6 \pi }+
\frac{\kappa ^2 \Omega }{24 \pi }  \label{M4}\\
-\frac{1}{2\pi} \VEV{J^{(4)r}_{ttt}} &=& \frac{7 \kappa ^4}{1920 \pi }+
\frac{ \Omega ^2 \kappa ^2}{16 \pi }+\frac{ \Omega ^4}{8 \pi }\0
\ee
where we have used $\VEV{f'(r_+)}_h = 2 \kappa$, $\VEV{T}_h=-\frac{\kappa^2}{2}$, 
$A_t(r) = \frac{r}{r^2+a^2}$, $\Omega=A_t(r_+)=\VEV{A_t}_h$. These results agree with
formula (\ref{Fn}) after the replacement $A_t\to m\, A_t$ (see the comment before
eq.(\ref{Ju1}). We checked the agreement up to spin 8 current 
\VEV{J^{(8)r}_{t\ldots t}}. The relevant expressions for  
 $\VEV{j^{(i)}_{u\ldots u}}_h$ for $i=1\ldots 8$ are given in the Appendix~B.

\section{Higher spin currents and trace anomalies}

Each of these higher spin currents carries to infinity its own component of
the Hawking radiation. Just in the same way as in the action
the metric is a source for the energy--momentum tensor, these new
(covariant) currents
will have in the effective action suitable sources, with  the appropriate
indices and symmetries. In \cite{BC} they were represented by
background fields $B^{(s)}_{\mu_1\ldots\mu_s}$ (which will be eventually set to zero). 
So we have
\be
 J^{(s)}_{\mu_1\ldots\mu_s}= \frac 1{\sqrt{g}} \frac {\delta}{\delta
B^{(s)\mu_1\ldots\mu_s}} S\label{Js}
\ee
We assume that all $J^{(s)}_{\mu_1\ldots\mu_s}$ are maximally symmetric and classically
traceless.

In addition to the series of $B^{(s)}$ fields, there must be other background fields
with the same characteristics (i.e. maximally symmetric and asymptotically trivial).
Their function is to explain the presence of the additional covariant terms
in the conservation equations of the higher currents (to be specific, the terms at
the RHS of (\ref{nablaJ3},\ref{nablaJ4},...). Let us call these additional fields
$C^{(s)},D^{(s)},...$. As an example let us consider the conservation of
$J^{(3)}$.
\be
\nabla^\mu J^{(3)}_{\mu\nu\lambda} = 2 F_\nu^\rho J^{(2)}_{\rho\lambda} -
\frac 16 (\nabla_\nu R) J^{(1)}_\lambda\label{nablaJ3cov}
\ee
where symmetrization over the indices $\nu$ and $\lambda$ is understood
in the RHS.
The LHS is due to assumed invariance of the effective action under
\be
\delta_\xi B^{(3)}_{\mu_1\mu_2\mu_3}=
\nabla_{\mu_1} \xi_{\mu_2\mu_3} + { cycl.}\label{deltaBxi}
\ee
where $\xi$ is a symmetric traceless tensor and ${ cycl}$
denotes cyclic permutations of the indices.
In order to explain the presence of the RHS terms, we assume
that there exist, in the effective
action, another background potentials $C^{(3)}$, coupled
to the two terms in the RHS of (\ref{nablaJ3cov}), which transforms like
\be
\delta_\xi C^{(3)}_{\mu\nu}=
\xi_{\mu\nu}, \ \label{deltaCxi}
\ee
while all the other fields in the game are invariant under $\xi$
transformations. These fields must have transformation properties that guarantee
the invariance of the terms they are involved in.

In an analogous way we can deal with the other conservation laws. We remark that
the transformations of the $C^{(s)}$ potentials are intrinsically Abelian.
Unfortunately we do not know how to derive  the
transformation (\ref{deltaCxi}) from first principles. But we can use
consistency to conclude that these two equations represent the only possibility.
For, although, in order to account for the $J^{(3)}$ conservation law, one
can envisage possible (non--Abelian) transformations one must check
that these transformations form a Lie algebra. Such a condition
strongly restricts the form of the transformations and, consequently, of
the effective action\footnote{The presence of the terms in the RHS of
(\ref{nablaJ3cov}) could be
formally explained by different transformation laws of the other fields.
In particular for the third order current such terms could be explained by
$$
\delta_\tau g_{\mu\nu}\sim \tau_\mu^\lambda F_{\lambda\nu},\quad\quad
\delta_\tau A_\mu \sim \tau_{\mu}^\lambda \nabla_\lambda R
$$
But these are not good symmetry transformations, for they do not form an algebra.
It is easy to see it for instance by promoting $\xi$ to anticommuting parameters
and verifying that such transformation are not nilpotent.}. One can indeed
verify that the higher potentials transformation laws
are so strongly restricted that it is generically impossible to avoid the conclusion
that they must be Abelian (see also II). Under these circumstances
(\ref{deltaCxi}) represents the generic case for higher spin quantities.
The presence of these additional background fields, which were not considered in I and II,
may complicate the anomaly analysis. However, to simplify it, one can remark that
these potentials can increase the number of cocycles only if they explicitly appear
in the cocycles themselves. Since eventually these potentials are set to zero,
the corresponding cocycles vanish. As a consequence they cannot give rise to 
the anomalies we are interested in and their study is of academical interest.
For this reason, for the sake of simplicity, we choose to dispense from it. Thus 
henceforth we will ignore the additional potentials.

This said we can now analyze the problem of the existence of trace anomalies
in higher spin currents with cohomological methods. With respect to I the
analysis is complicated by the presence of the
electromagnetic field. 
Of course the electromagnetic field gives rise to the gauge anomaly in the covariant
derivative of the $J^{(1)}$ current, see (\ref{nablaJ1}). The latter is induced by 
the gauge transformation 
$\delta_\lambda A_\mu=  \partial_\mu \lambda$ and this is all we need to say about
this anomaly.

With these premises, we want to show that the conclusion of I on the absence
of trace anomalies in the higher spin currents holds under the present
conditions. Let us recall first the setting of I for this type of analysis, 
\cite{Bonora83,Bonora84,Bonora85}.
Let us start from the analysis of $J^{(3)}$. Setting $B^{(3)}_{\mu\nu\lambda}=
B_{\mu\nu\lambda}$ 
the Weyl transformation of the various field involved are (see I for a comparison)
\be
\delta_\sigma g_{\mu\nu}&=& 2 \sigma\, g_{\mu\nu}\0\\
\delta_\sigma B_{\mu\nu\lambda}&=&x\, \sigma B_{\mu\nu\lambda}\label{deltasigma}\\
\delta_\sigma A_\mu&=&0
\ee
which induces the trace of the energy--momentum tensor,
and
\be
\delta_{\tau}  g_{\mu\nu}&=&0 \0\\
\delta_{\tau}  B_{\mu\nu\lambda}&=& \tau_\mu g_{\nu\lambda}+{\rm cycl}\label{deltatau}\\
\delta_\tau A_\mu&=&0
\ee
which induces the trace of $J^{(3)}$. Moreover, for consistency with (\ref{deltasigma})
we must have
\be
\delta_\sigma \tau_\mu= (x-2) \sigma \tau_\mu\label{deltasigmatau}
\ee
where $x$ is an arbitrary number. 

A comment on these transformations is in order. They are determined as follows:
they must be expressed in terms of 
symmetry parameters and of the basic background fields $g_{\mu\nu}$ and $A_\mu$
and nothing else; they must form a Lie algebra, as was mentioned above, and they must
leave unchanged the terms in the effective action, in particular the terms involving
the matter fields. The transformations are then dictated by the canonical dimensions 
of the various fields. The fields $B^{(s)}$ and $C^{(s)}$ have dimension $2-s$ and $1-s$,
respectively.  

We must now repeat the analysis we have done in I. We promote $\sigma$ and $\tau_\mu$ to
anticommuting fields so that 
\be
\delta_\sigma^2=0,\quad\quad \delta_\tau^2=0,\quad\quad \delta_\sigma\, 
\delta_\tau+ \delta_\tau\, \delta_\sigma=0\0
\ee
Integrated anomalies are defined by 
\be
\delta_\sigma \Gamma^{(1)} = \hbar\, \Delta_\sigma,
\quad\quad \delta_\tau \Gamma^{(1)} = \hbar\, \Delta_\tau,\label{cocycles}
\ee
where $\Gamma^{(1)}$ is the one--loop quantum action and 
$\Delta_\sigma,\Delta_\tau$ are
local functional linear in $\sigma$ and $\tau$, respectively. The 
unintegrated anomalies, i.e. the traces $T_\mu^\mu$ and 
$J^{(3)\mu}{}_{\mu\la}$ are obtained
by functionally differentiating with respect to 
$\sigma$ and $\tau_{\la}$, respectively.

By applying $\delta_\sigma,\delta_\tau$ to the eqs.\refb{cocycles}, we see 
that candidates for anomalies  $\Delta_\sigma$ and $\Delta_\tau$
must satisfy the consistency conditions
\be\label{condss}
\delta_\sigma\, \Delta_\sigma  = 0,\quad 
\delta_\tau\, \Delta_\sigma + \delta_\sigma \,\Delta_\tau = 0,\quad
\delta_\tau\,\Delta_\tau = 0
\ee 

Once we  have determined these cocycles we have to make sure that they are true 
anomalies, that is that they are nontrivial. In other words there must 
not exist local counterterm $C$ in the action such that 
\begin{eqnarray} \label{condtriv1}
	\Delta_\sigma &=& \delta_\sigma\, \int d^2x \sqrt{-g}  \,{\cal C}\\
\label{condtriv2}
\Delta_\tau &=&  \delta_\tau\, \int d^2x\, \sqrt{-g}\,  {\cal C}
\end{eqnarray}
If such a $C$ existed we could redefine the quantum action by subtracting these 
counterterms and get rid of the (trivial) anomalies.

Let us consider now the problem of the trace $J^{(3)\mu}{}_{\mu\la}$. We could repeat
the complete analysis of I, but there is a shortcut due to the simple form of the 
transformations (\ref{deltatau}). Suppose we find cocycle $\Delta^{(3)}_\tau$
\be
\Delta^{(3)}_\tau =  \int d^2x \sqrt{-g}\, \tau^{\mu} \,I^{(3)}_{\mu}\label{defDt3}
\ee
where $I^{(3)}_{\mu}$ is a canonical dimension 3 tensor made of the metric, the gauge field and their
derivatives, such as $\nabla_\mu R$ or $\nabla_\nu F_{\mu}{}^\nu$,
or even a non--gauge--invariant tensor such as $A_\mu R$. Then it is immediate to
write down a counterterm
\be
{\cal C}^{(3)}\sim B^{\mu\la}_{\la} \,I^{(3)}_{\mu}\label{countC}
\ee
which cancels (\ref{defDt3})\footnote{Of course the variation of (\ref{countC}) 
with respect to $\sigma$ gives rise to a trivial $\Delta_\sigma$ cocycle, but this cocycle
depends on the field $B$, which vanishes when we select the physical background. 
On the other hand, we  have shown in I, in an analogous case, that such cocycles can be 
consistently eliminated even without resorting to the vanishing of $B$.}. 

As for the trace $J^{(4)\mu}{}_{\nu\la\rho}$ we can proceed in analogy to 
$J^{(3)\mu}{}_{\mu\la}$. Setting
$B^{(4)}_{\mu\nu\la\rho}\equiv B_{\mu\nu\la\rho} $,
the relevant Weyl 
transformations are as follows.  
The variation $\delta_\tau$ acts only on $B_{\mu \nu\lambda\rho}$
\begin{equation}
\delta_\tau B_{\mu \nu \lambda\rho} = 
g_{\mu \nu}\,\tau_{ \lambda\rho} +{\rm cycl}
 \label{tautransf}
\end{equation}
and the other fields remain unchanged while the
variation with respect to the ordinary Weyl parameter $\sigma$
are
\begin{eqnarray} 
&& \delta_\sigma g_{\mu \nu}  = 2 \,\sigma\, g_{\mu \nu}  \\
&& \delta_\sigma \tau_{\mu \nu}  = (x-2)\,\sigma\, \tau_{\mu \nu} \\
&& \delta_\sigma B_{\mu \nu\lambda \rho} = x\,\sigma\, 
B_{\mu \nu\lambda\rho}
\label{sigmatransf}
\end{eqnarray}
where, again, $x$ is an arbitrary number.
Now we can repeat the previous argument. Let a cocycle have the form
\begin{equation}
\Delta^{(4)}_\tau =  \int d^2x \sqrt{-g}\, \tau^{\mu\nu} \,I^{(4)}_{\mu\nu}\label{defDt4}
\end{equation}
where $I^{(4)}_{\mu}$ is a dimension 4 tensor made out of the metric, the gauge field and 
their derivatives, such as $\nabla_\mu\nabla_\nu R$. The counterterm
\be
{\cal C}^{(4)}\sim B^{\mu\nu\la}_{\la} \,I^{(4)}_{\mu\nu}\label{countC4}
\ee
cancels (\ref{defDt4}).

It is not hard to generalize this conclusion to higher spin currents.
We believe these results together with those of I and II are evidence enough that
anomalies may not arise in the higher spin currents under any condition.

\section{Current normalization and $W_{1+\infty}$ algebra}

It is evident that our being able to describe the higher moments of the Hawking
radiation is related to the transformation properties of the holomorphic higher
spin currents. Even in the case of the energy momentum tensor, the Hawking
flux is related to Weyl or Diff anomalies only in the sense that the latter
determine the relation between the covariant and holomorphic part of the
energy--momentum tensor (see our discussion in II). For higher spin currents,
as we have seen, there are no links with anomalies simply because anomalies
cannot exist in the conservation laws of these currents. This much seems
definitely clear. There are however
other aspects of the problem which have remained implicit and are crucial
in order to understand the central role of the $W_{1+\infty}$ algebra.
In this section we would like to discuss these aspects.

Let us start from the remark that in formula (\ref{XF1}) the summation over $k$
does not affect the crucial term $\frac{\kappa^{s}}{s} (1 - 2^{-(s-1)}) B_s$
except for an overall multiplicative factor. This means that, had we used
each one of the currents
\be
j_{z\ldots z}^{(s,k)}(z)=\partial_z^{s-k} \Psi^{\dagger}(z) \, \partial_z^{k-1}
\Psi(z) :\label{jsk}
\ee
instead of (\ref{js}), we would have obtained (up to normalization) the same final
result for the moments of the Hawking radiation\footnote{This, by the way, explains why
the authors of \cite{IMU4,IMU5} obtained the same predictions as in I and II
for the higher moments of the Hawking radiation, using unnormalized currents
and without invoking a $W_\infty$ structure.}. This seems at first to deprive of any interest
the role of the $W_{1+\infty}$ algebra, but the case is just the opposite.
Using the currents $j_{z\ldots z}^{(s,k)}(z)$ we have two enormous disadvantages.

The first is that we do not have any means of normalizing these currents, thus
rendering the results obtained by their means devoid of any predictive value.
The $W_{1+\infty}$ algebra structure tells us how to normalize the currents in such a way
as to get an algebra. There remain only two constants to be fixed $\lambda$ and
$q$. The first is fixed in such a way as to get the right transformation laws
(OPE) of the energy--momentum tensor, the second is fixed by the U(1) algebra
of $j^{(1)}$. Once these two constants are fixed the normalization for all the
higher spin currents is uniquely determined and in agreement with the thermal
spectrum of the Hawking radiation.

The second disadvantage of using not $W_{1+\infty}$ currents, such as
$j_{z\ldots z}^{(s,k)}(z)$, is the appearance of anomalies in their traces
or in the conservation laws of their covariant version. This was shown in
a very explicit way in \cite{IMU4}. As we have shown, these anomalies are
cohomologically trivial and can be eliminated by suitable redefinitions or
subtractions. As a result one ends up with the currents (\ref{js}) and
their $W_{1+\infty}$ algebra. In other words the $W_{1+\infty}$ algebra
is the appropriate structure underlying the thermal spectrum of the Hawking
radiation. This result seems to imply that the two--dimensional physics
around the horizon is characterized by a symmetry larger than the Virasoro
algebra, such as a $W_\infty$ or $W_{1+\infty}$ algebra.

\acknowledgments

We would like to thank P.~Dominis~Prester for useful discussions. 
M.C. would like to thank SISSA for hospitality and The National Foundation for Science, Higher Education
and Technological Development of the Republic of Croatia (NZZ) for
financial support. I.S., M.C.\ and S.P.\ would like to acknowledge support 
by the Croatian Ministry of Science, Education and Sport under the 
contract no.119-0982930-1016. L.B. would like to thank the GGI in Florence
for hospitality during this research.

\appendix
\section{Spin connection}

With reference to section 2 we list here the spin connection coefficients 
$\omega^a{}_{bc} = e_c{}^\mu \omega^a{}_{b\mu} = e_c{}^\mu e^a{}_\nu \nabla_\mu e_b{}^\nu$
for the Kerr metric \refb{4metric} and the vierbein \refb{4bein}:
\be
\omega^0{}_{10} &=& \frac{(r-M) \Sigma (r,\theta )-r \Delta (r)}{\sqrt{\Delta (r)} \Sigma (r,\theta )^{3/2}}\0\\
\omega^0{}_{20} &=& \omega^1{}_{21} = -\frac{a^2 \cos \theta  \sin \theta }{\Sigma (r,\theta )^{3/2}}\0\\
\omega^1{}_{30} &=& \omega^0{}_{31} = \omega^0{}_{13} = -\frac{a r \sin \theta }{\Sigma (r,\theta )^{3/2}}\0\\
\omega^2{}_{30} &=& -\omega^0{}_{32} = \omega^0{}_{23} = -\frac{a \cos \theta  \sqrt{\Delta (r)}}{\Sigma (r,\theta )^{3/2}}\0\\
\omega^1{}_{22} &=& \omega^1{}_{33} = -\frac{r \sqrt{\Delta (r)}}{\Sigma (r,\theta )^{3/2}}\0\\
\omega^2{}_{33} &=& -\frac{(2 M r+\Delta (r)) \cos \theta }{ \Sigma (r,\theta )^{3/2} \sin \theta }\0
\ee
Apart from these coefficients and those related by using $\omega_{abc} = 
-\omega_{bac}$, where $\omega_{abc} = \eta_{ad} \omega^d{}_{bc}$, all other coefficients 
are zero. Christoffel symbols for the metric \refb{4metric} can be found in Appendix~D 
of \cite{Frolov}.

\section{Fifth order current}

In this Appendix we write down the explict expression for the fifth order current
and the values of the holomorphic currents at the horizon, up to order eight.

The current $J^{(5)}_{uuuuu}$ is as follows:
\be
J^{(5)}_{uuuuu} &=& 
\hbar  \left( \frac{32 A_u^5}{5}
-\frac{104 T A_u^3}{21}-\frac{16}{7} \left(\nabla _u^2A_u\right) A_u^2+\frac{27 T^2 A_u}{70}+\frac{24}{7} \left(\nabla _uA_u\right)^2 A_u
\right. \CR && \qquad \left. 
+\frac{1}{35} \left(\nabla _u^2T\right) A_u-\frac{1}{7} \left(\nabla _uA_u\right) \left(\nabla _uT\right)
+\frac{2}{21} T \left(\nabla _u^2A_u\right)\right)
\CR &&
-16 J^{(1)}_{u} A_u^4-32 J^{(2)}_{uu} A_u^3+\frac{8}{7} \left(\nabla _u^2J^{(1)}_{u}\right) A_u^2+
\frac{52}{7} T J^{(1)}_{u} A_u^2-24 J^{(3)}_{uuu} A_u^2
\CR &&
-\frac{24}{7} \left(\nabla _uA_u\right) \left(\nabla _uJ^{(1)}_{u}\right)
A_u+\frac{12}{35} \left(\nabla _u^2J^{(2)}_{uu}\right) A_u+\frac{16}{7} 
\left(\nabla _u^2A_u\right) J^{(1)}_{u} A_u
\CR &&
+\frac{52}{7} T J^{(2)}_{uu} A_u-8 J^{(4)}_{uuuu} A_u 
+\frac{1}{14} \left(\nabla _uT\right) \left(\nabla _uJ^{(1)}_{u}\right)-\frac{12}{7} \left(\nabla _uA_u\right) \left(\nabla _uJ^{(2)}_{uu}\right)
\CR &&
-\frac{1}{21} T \left(\nabla _u^2J^{(1)}_{u}\right)-\frac{27 T^2 J^{(1)}_{u}}{140}-\frac{12}{7} \left(\nabla _uA_u\right)^2 J^{(1)}_{u}
\CR &&
-\frac{1}{70} \left(\nabla _u^2T\right) J^{(1)}_{u}+\frac{8}{7} \left(\nabla _u^2A_u\right) J^{(2)}_{uu}+\frac{13 T J^{(3)}_{uuu}}{7}+j^{(5)}_{uuuuu}
\ee
For simplicity we omit the explicit expressions of $J^{(6)}_{u\ldots u}$, 
$J^{(7)}_{u\ldots u}$, $J^{(8)}_{u\ldots u}$. Next we list the results 
for $j^{(i)}_{u\ldots u}$ at the horizon, obtained using the  
condition that $J^{(i)}_{\u\ldots \u}$, in Kruskal coordinates, be regular. We understand
$\la=\hbar=1$:
\be
\VEV{j^{(1)}_{u}}_h &=& -\VEV{A_t}_h \0\\
 \VEV{j^{(2)}_{uu}}_h &=& -\frac{1}{2}\VEV{A_t}_h^2 + \frac{1}{12}\VEV{T}_h \0\\
 \VEV{j^{(3)}_{uuu}}_h &=& -\frac{1}{3}\VEV{A_t}_h^3 + \frac{1}{6}\VEV{T}_h 
\VEV{A_t}_h \0\\
 \VEV{j^{(4)}_{uuuu}}_h &=& -\frac{1}{4}\VEV{A_t}_h^4+\frac{1}{4}\VEV{T}_h 
\VEV{A_t}_h^2-\frac{7}{240}\VEV{T}_h^2 \0\\
 \VEV{j^{(5)}_{uuuuu}}_h &=& -\frac{1}{5}\VEV{A_t}_h^5+\frac{1}{3}\VEV{T}_h 
\VEV{A_t}_h^3-\frac{7}{60}\VEV{T}_h^2 \VEV{A_t}_h \0\\
 \VEV{j^{(6)}_{uuuuuu}}_h &=& -\frac{1}{6}\VEV{A_t}_h^6+\frac{5 }{12}\VEV{T}_h 
\VEV{A_t}_h^4-\frac{7}{24}\VEV{T}_h^2 \VEV{A_t}_h^2+\frac{31 }{1008}\VEV{T}_h^3 \0\\
 \VEV{j^{(7)}_{uuuuuuu}}_h &=& -\frac{1}{7}\VEV{A_t}_h^7+\frac{1}{2}\VEV{T}_h 
\VEV{A_t}_h^5-\frac{7}{12}\VEV{T}_h^2 \VEV{A_t}_h^3+\frac{31}{168} \VEV{T}_h^3 \VEV{A_t}_h \0\\
 \VEV{j^{(8)}_{uuuuuuuu}}_h &=& -\frac{1}{8}\VEV{A_t}_h^8+\frac{7}{12}\VEV{T}_h 
\VEV{A_t}_h^6-\frac{49}{48}\VEV{T}_h^2 \VEV{A_t}_h^4+\frac{31}{48} \VEV{T}_h^3 
\VEV{A_t}_h^2-\frac{127}{1920}\VEV{T}_h^4\0
\ee

\end{document}